\documentstyle[12pt]{article}

\begin{document}

\title{Gravitation as a Supersymmetric \\Gauge Theory}
\author{Roh S. Tung\\
         {\it Enrico Fermi Institute and Department of Physics}\\
         {\it University of Chicago}\\
         {\it 5640 South Ellis Avenue}\\
         {\it Chicago, Illinois 60637-1433, USA}\\
         {\it rohtung@midway.uchicago.edu}}
\maketitle

\begin{abstract}

We propose a gauge theory of gravitation.
The gauge potential is a connection of the Super SL(2,C) group.
A MacDowell-Mansouri type of action is proposed where
the action is quadratic in the Super SL(2,C) curvature
and depends purely on gauge connection.
By breaking the symmetry of the Super SL(2,C)
topological gauge theory to SL(2,C),
a spinor metric is naturally defined.
With an auxiliary anti-commuting spinor field,
the theory is reduced to general relativity.
The Hamiltonian variables are related to
the ones given by Ashtekar.
The auxiliary spinor field
plays the role of Witten spinor
in the positive energy proof
for gravitation.
\end{abstract}
\newpage

It was Einstein's great insight that gravity is
a manifestation of space-time curvature.
In Riemannian geometry,
a space-time with curvature is described by the metric
and Einstein therefore used it to describe gravity.

The prime principle of theoretical physics is,
however, to begin with the action.
It gives consistent interacting field equations
with the desired conserved quantities.
Moreover all other basic interactions have
a variational foundation.
An obvious candidate for the gravitational variables for the action
is the metric (or the tetrad).
However, nature may prefer other candidates.
In view of the success of gauge theory,
the question of whether gravity can be written in terms of
a pure connection theory has been addressed \cite{Gauge}.

Important progress was made with the Ashtekar's New Variables \cite{A},
in which a self-dual connection is the basic Hamiltonian variable.
The momentum conjugate to the self-dual connection
is the ``triad'' made from a self-dual 2-form.
Cappovilla, Dell and Jacobson then proposed a purely
connection action for the New Variables \cite{CDJ}.

Recently, a quadratic spinor Lagrangian
for general relativity was proposed,
where a Dirac spinor 1-form field \cite{NT}
(or two 2-component spinor 1-form fields \cite{TJ}) was considered
as the variational variable.
(Another 2-form version of the Lagrangian was also discovered
by Robinson \cite{R1}).
The spacetime metric is represented by a
quadratic combination of the spinor
1-form field.
This provides a succinct representation for general
relativity, and provides a way to represent
the gravitational energy-momentum \cite{NT}.
With a reality condition satisfied, the quadratic spinor Lagrangian
is equivalent to
the teleparallel Lagrangian for general relativity \cite{TN99}.
Robinson discussed the theory using a Lie algebra \cite{R2}.

As the variable has half helicity,
one may ask if it can be
an anti-commuting field.
Jacobson suggested using a single anti-commuting spinor 1-form \cite{J};
Robinson provided a formulation with two anti-commuting spinor
1-forms \cite{R1}.
Recently we learned that another similar
anti-commuting approach was used
long ago in the context of supergravity \cite{BM}.

The purpose of this letter is to propose a gauge theory
along these lines.
We start from a gauge theory of
the Super SL(2,C) group, giving a
topological gauge theory of the
MacDowell-Mansouri type.
By breaking this supersymmetry into the SL(2,C) symmetry,
we get two equivalent chiral actions.
We show that, up to a nilpotent metric,
the theory is equivalent to general relativity.

Let us start with a Super SL(2,C) algebra
(with three SL(2,C) generators $M_{00},M_{01}=M_{10},M_{11}$
and two supersymmetric generators $Q_0,Q_1$):

\begin{eqnarray}
\left[ M_{AB}, M_{CD} \right] &=&
      \epsilon_{C(A} M_{B)D}
     +\epsilon_{D(A} M_{B)C} , \\
\left[ M_{AB}, Q_C \right] &=& \epsilon_{C(A} Q_{B)} ,\\
\{Q_A, Q_B\} &=& 2 M_{AB} ,
\end{eqnarray}
where $\epsilon_{C(A} M_{B)D}
=\textstyle{1\over2} (\epsilon_{CA} M_{BD}
+\epsilon_{CB} M_{AD})$ and 
$\epsilon_{C(A} Q_{B)}
=\textstyle{1\over2} (\epsilon_{CA} Q_{B}
+\epsilon_{CB} Q_{A})$.

The upper-case Latin letters $A,B,...=0,1$ denote two component spinor
indices, which are raised and lowered with the constant
symplectic spinors
$\epsilon_{AB}=-\epsilon_{BA}$
together with its inverse and their conjugates
according to the conventions $\epsilon_{01}=\epsilon^{01}=+1$,
$\lambda^A:=\epsilon^{AB}\lambda_B$,
$\mu_B:=\mu^A\epsilon_{AB}$ \cite{PR}.
Lowercase Latin letters
$p,q,...$ denote the Super SL(2,C) group indices,
$a,b,c,...=0,1,2,3$ denote the SO(3,1) indices .

The Super SL(2,C) group
is isomorphic to the complex extension of
OSp(1,2). It is a {\it simple} super Lie group
and has a nondegenerate Killing form \cite{DeWitt}.
The Cartan-Killing metric
$\eta_{pq}={\rm{diag}}(\eta_{(AB)(MN)},\eta_{AB})$ is given by
\begin{eqnarray}
\eta_{(AB)(MN)}&=&
\textstyle{1\over2}
(\epsilon_{AM}\epsilon_{BN}+\epsilon_{AN}\epsilon_{BM}), \\
\eta_{AB}&=& -2 \epsilon_{AB} .
\end{eqnarray}

To {\it gauge} this Super SL(2,C) group, we associate
to each generator $T_p=\{M_{AB}, Q_A\}$ a 1-form field
$A^p=\{\omega^{AB},\varphi^{A}\}$, and form a
super Lie algebra valued connection 1-form,
\begin{equation}
A=A^p T_p=\omega^{AB} M_{AB} +\varphi^A Q_A ,
\label{1}
\end{equation}
where $\omega^{AB}$ is the SL(2,C) connection 1-form and
$\varphi^A$ is an anti-commuting spinor valued 1-form.
(We shall use ${\cal D}$ for the Super SL(2,C) covariant derivative
and $D$ for the SL(2,C) covariant derivative.)

The curvature is given by
$F=dA + \textstyle{1\over2}[A, A]=
dA+\textstyle{1\over2}A^p\wedge A^q \otimes [T_p, T_q]$.
Given the Super SL(2,C) connection $A$
defined in equation (\ref{1}), the curvature
($F=F(M)^{AB} M_{AB}+F(Q)^A Q_A$) contains
a bosonic part associated with $M_{AB}$,
\begin{equation}
F(M)^{AB}=d\omega^{AB}+\omega^{AC}\wedge \omega_C{}^B
+ \varphi^A\wedge\varphi^B ;
\end{equation}
and a fermionic part associated with $Q_A$ ,
\begin{equation}
F(Q)^A=d\varphi^A+\omega^{AB}\wedge \varphi_B .
\end{equation}

A simple {\em spinor} action, quadratic in the curvature,
using this Super SL(2,C) connection $A$ is
\begin{eqnarray}
{\cal S}_{\rm{T}}[A^p]&=&\int\, F^p\wedge F^q \,\eta_{pq} \nonumber\\
&=&\int\, F(M)^{AB}\wedge F(M)_{AB} \nonumber\\
&&\qquad + 2 F(Q)^{A}\wedge F(Q)_{A} ,
\end{eqnarray}
where $\eta_{pq}$ is the Cartan-Killing metric of the Super SL(2,C) group,
${\cal D}\eta_{pq}=0$.
However, this action is a total differential.
Hence, similar to the work of MacDowell and Mansouri \cite{MM},
we need to choose another
spinor action which is SL(2,C) invariant,
thus breaking the
topological field theory of the Super SL(2,C) symmetry
into an SL(2,C) symmetry.
Let us choose $i_{pq}={\rm{diag}}(i_{(AB)(MN)},i_{AB})$ such that
\begin{eqnarray}
i_{(AB)(MN)}&=&\textstyle{1\over2}
\left(
\epsilon_{AM}\epsilon_{BN}
+\epsilon_{AN}\epsilon_{BM}
\right) , \\
i_{AB}&=& 0 .
\end{eqnarray}
The new spinor action is
\begin{eqnarray}
{\cal S}[A^p]&=&\int\, F^p\wedge F^q \, i_{pq} \nonumber\\
&=&\int\,  F(M)^{AB} \wedge F(M)_{AB} .
\end{eqnarray}
In terms of the SL(2,C) curvature
$R^{AB}=d\omega^{AB}+\omega^{AC}\wedge \omega_C{}^B$,
the spinor action ${\cal S}[A^p]$ is
\begin{equation}
{\cal S}[\omega^{AB},\varphi^A]
=\int\,   R^{AB}\wedge R_{AB}
+ 2 R_{AB}\wedge \varphi^A \wedge \varphi^B . \label{action}
\end{equation}
Note that, due to SL(2,C) Bianchi identity,
the first term is purely topological, while
the second term gives the dynamics.
(The $\varphi^4$-term vanishes.)
In the following discussion we consider only
the dynamical term.

The field equations are obtained by varying the
Lagrangian with respect to
the gauge potentials---the Super SL(2,C) connection.
With these gauge potentials fixed at the boundary, the field equations are
\begin{eqnarray}
&& R^{AB}\wedge\varphi_B=0 \qquad (DF(Q)^A=0),\label{fe1} \\
&&D(R^{AB}+ \varphi^A \wedge \varphi^B)=0 \qquad (DF(Q)^{AB}=0),\label{fe2}
\end{eqnarray}
where, becuase of the SL(2,C) Bianchi identidy
($DR^{AB}=0$), the second field equation (\ref{fe2}) is reduced to
$D(\varphi^A \wedge \varphi^B)=0$.

Another choice of the spinor action which is SL(2,C) invariant is that
$i'_{pq} = {\rm{diag}} (i'_{(AB)(MN)} , i'_{AB})$ where
\begin{eqnarray}
i'_{(AB)(MN)}&=&0 ,\\
i'_{AB}&=& -2 \epsilon_{AB} ,
\end{eqnarray}
then the new spinor action is
\begin{eqnarray}
{\cal S'}[A^p]&=&\int\, F^p\wedge F^q \, i'_{pq} \nonumber\\
&=&\int\, 2 F(Q)^{A} \wedge F(Q)_{A}  \nonumber\\
&=&\int\, 2 D\varphi^{A} \wedge D\varphi_{A} . \label{action1}
\end{eqnarray}
This action differs from the previous one by
a total differential, a surface term:
\begin{equation}
{\cal S}=-{\cal S'}+\oint 2D\varphi^A\wedge\varphi_A .
\label{ss}
\end{equation}
Thus they give the same field equations.

In order to make a connection between the internal space
of the Super SL(2,C) group with the structures on the four-manifold,
we follow a construction in Kerrick \cite{Kerrick}:
we observe that although ${\cal D}\eta_{pq}=0$,
${\cal D}i_{pq}$ does not vanish, thereby
providing
a {\it soldering form} from the tangent space of the
four-manifold to the internal space of Super SL(2,C).

Therefore,
when we break the symmetry from
a Super SL(2,C) topological field theory
${\cal S}_{\rm T}[A^p]$ into an SL(2,C) invariant ${\cal S}[A^p]$ ,
a {\em spinor metric} is {\em naturally} defined.
Using the fact that
\begin{equation}
{\cal D}i_{pq}=C^m{}_{pn} A^n i_{mq} ,
\end{equation}
and ${\cal D}\eta_{pq}=0$,
the {\it spinor metric} ${\cal G}$ is defined by
\begin{equation}
{\cal G}=\eta^{pm}\eta^{qn} {\cal D}i_{pq} \otimes {\cal D}i_{mn}
=\epsilon_{AB} \varphi^A \otimes \varphi^B.
\label{spinormetric}
\end{equation}
Because
${\cal D}\eta_{pq}={\cal D}i_{pq}+{\cal D}i'_{pq}=0$,
the {\it same} metric is defined when we choose the
action ${\cal S'}$ with $i_{pq}$ replaced by $i'_{pq}$.

For non-degenerate solutions
($\varphi^0\wedge\varphi^1\wedge
\overline\varphi^{0'}\wedge\overline\varphi^{1'}\ne0$)
in which
$\varphi^0, \varphi^1$ and their complex conjugates
$\overline\varphi^{0'},\overline\varphi^{1'}$
are linear independent,
the 2-form
$\varphi^{A}\wedge\varphi^{B}$
and their complex conjugates
$\varphi^{A'}\wedge\varphi^{B'}$
form a basis for the
six-dimensional space of 2-form.
We can form a 1-1 map such that:
\begin{eqnarray}
\varphi^{0}\wedge\varphi^{0} &\mapsto& \theta^{00'}\wedge\theta^{01'}, \\
\varphi^{0}\wedge\varphi^{1} &\mapsto&
\textstyle{1\over2} (\theta^{00'}\wedge\theta^{11'}
-\theta^{01'}\wedge\theta^{10'}),\\
\varphi^{1}\wedge\varphi^{1} &\mapsto& \theta^{10'}\wedge\theta^{11'} ,
\end{eqnarray}
and their complex conjugates
\begin{eqnarray}
\overline\varphi^{0'}\wedge\overline\varphi^{0'} &\mapsto&
\theta^{00'}\wedge\theta^{10'}, \\
\overline\varphi^{0'}\wedge\overline\varphi^{1'} &\mapsto&
\textstyle{1\over2} (\theta^{00'}\wedge\theta^{11'}
+\theta^{01'}\wedge\theta^{10'}),\\
\overline\varphi^{1'}\wedge\overline\varphi^{1'} &\mapsto&
\theta^{01'}\wedge\theta^{11'}.
\end{eqnarray}
This can be established by introducing
an anti-commuting spinor field $\xi^{A'}$ such that
$\bar\kappa=\epsilon_{A'B'}\xi^{A'}\xi^{B'}$ is a nilpotent constant.
(This can be constructed in terms of an anti-commuting dyad where
$\xi^{0'}=o^{0'}$, $\xi^{1'}=\iota^{1'}$
are odd constant spinors. They are spin basis with
$o^{A'}o_{A'}=\iota^{A'}\iota_{A'}=0$,
$\iota^{0'}=o^{1'}=0$, $o_{A'}\iota^{A'}=2 o^{0'}\iota^{1'}=\bar\kappa$,
where $\bar\kappa$ is a nilpotent constant even element of
the Grassmann algebra, see also Robinson\cite{R1}.)

Now we relate the field $\varphi^A$
to the tetrad field $\theta^{AA'}$ by
$\theta^{AA'}=\varphi^A\xi^{A'}$,
thus
\begin{equation}
\bar\kappa \, \varphi^A\wedge\varphi^B
=\theta^{(AA'}\wedge\theta^{B)}{}_{A'}.
\end{equation}
The {\em spacetime} metric is then defined
by multiplying the {\em spinor} metric (\ref{spinormetric}) by
a nilpotent constant $\bar\kappa$,
\begin{equation}
g=\bar\kappa \otimes {\cal G}
 =\epsilon_{AB}\epsilon_{A'B'}\theta^{AA'}\otimes\theta^{BB'} ,
\end{equation}
which is
the usual tetrad expression for spacetime metric.

Multiplying the spinor action (\ref{action}) by $\bar\kappa$, we get
\begin{eqnarray}
S[A^p,\bar\kappa]&=&\bar\kappa{\cal S}[A^p] \nonumber\\
&=& \int\, 2 R_{AB} \wedge \theta^{AA'}\wedge\theta^{B}{}_{A'} ,
\end{eqnarray}
the usual chiral action for general relativity .

By varying the action $S$ with respect to
$\theta^{AA'}=\varphi^{A}\xi^{A'}$,
we obtained the chiral Einstein equation,
\begin{equation}
R^{AB}\wedge\varphi_B\xi_{B'}= R^{AB}\wedge\theta_{BB'}=0 .
\end{equation}
Varying $\omega^{AB}$, we obtained
\begin{equation}
\bar\kappa D(\varphi^A\wedge\varphi^B)
=D(\theta^{(AC'}\wedge\theta^{B)}{}_{C'})=0,
\end{equation}
which is the torsion free equation.

If we start with the action ${\cal S'}$,
and multiply the spinor action by $\bar\kappa$,
by using (\ref{ss}), we get
\begin{eqnarray}
&& S'[A^p,\bar\kappa]=\bar\kappa{\cal S'}[A^p] \nonumber\\
&=& \int\, \bar\kappa
\left( -2 R_{AB}\wedge \varphi^A\wedge\varphi^B
+d(2 D\varphi^A\wedge \varphi_A) \right) \nonumber\\
&=& \int\,
-2 R_{AB} \wedge \theta^{AA'}\wedge\theta^{B}{}_{A'}
+d(2 \bar\kappa D\theta^{AA'}\wedge \theta_{AA'}) ,
\end{eqnarray}
where we used $d\bar\kappa=0$.
It reduces to the same action as $S$ up to a sign and
a total differential, and thus gives the same field equations.

The Hamiltonian can be constructed from either
the action (\ref{action}) or (\ref{action1})
by making a space-time decomposition
\begin{equation}
S= \bar\kappa
 \int dt\int d^3x\,\, \dot{A}^p_i \pi^i_p
-{\cal H} .
\end{equation}
For the action (\ref{action}), the canonical
momentum are:
\begin{eqnarray}
\pi^{iA}=0, \qquad
\pi^{iAB}=\epsilon^{ijk}\varphi^A_j\varphi^B_k .
\end{eqnarray}
This yields the Hamiltonian density
\begin{equation}
{\cal H}=
 \omega^{AB}_0 {\cal H}_{AB}+
 \varphi^A_0 {\cal H}_A ,
\label{h0}
\end{equation}
where
\begin{equation}
{\cal H}_{AB}=D_i\pi^i_{AB}=0 ,
\end{equation}
and
\begin{equation}
{\cal H}_A=\epsilon^{ijk} R_{ijAB}\varphi^B_k=0 ,
\end{equation}
are the constraints.

The canonical variables are related to the ones used
in Ashtekar \cite{A} by
\begin{eqnarray}
&&\tilde\sigma^{iMN}=\bar\kappa\pi^{iMN}=
\bar\kappa\epsilon^{ijk}\varphi^M_j\varphi^N_k , \\
&&q_{ij}=\bar\kappa\epsilon_{AB}\varphi^A_i\varphi^B_j ,
\end{eqnarray}
where $\tilde\sigma^{iMN}$ is a densitized triad and $q_{ij}$
is the 3 metric.

For the action (\ref{action1}), the canonical momentum are:
\begin{eqnarray}
\pi^{iA}=\epsilon^{ijk}D_j\varphi^A_k, \qquad
\pi^{iAB}=0 ,
\end{eqnarray}
yielding a Hamiltonian density
\begin{equation}
{\cal H}'=
 \omega^{AB}_0 {\cal H}'_{AB}+
 \varphi^A_0 {\cal H}'_A
  +\partial_i (\varphi^A_0\pi^i_A) ,
\label{h}
\end{equation}
where
\begin{equation}
{\cal H}'_{AB}=\varphi_{iB}\pi^i_A=0 ,
\end{equation}
and
\begin{equation}
{\cal H}'_A=D_i\pi^i_A=0 ,
\end{equation}
are the constraints.

The Hamiltonian (\ref{h}) differs from (\ref{h0}) by a
total differential.
It has asymptotically flat fall off of $O(1/r^4)$, and
the variation of the Hamiltonian will have an $O(1/r^3)$ boundary term
which vanishes asymptotically.
Therefore there is no need of adding a further boundary
term for the Hamiltonian (\ref{h}) \cite{RT}.
With the field equations satisfied,
the Hamiltonian (\ref{h}) reduces to an exact differential.
Integration yields an integral over a 2-surface
\begin{equation}
 E=\bar\kappa \int {\cal H}' \mid_{\rm S}
   =\bar\kappa \oint \varphi^A_0 D\varphi_A ,
 \label{energy}
\end{equation}
which determines the energy within a 2-surface.

This is comparable with the usual spinorial quasilocal energy
constructions
\begin{equation}
 E_w =\oint
{\xi}_{A} D\xi_{A'}\wedge \theta^{AA'} ,
\end{equation}
with $\theta^{AA'}=\varphi^A\xi^{A'}$,
it reduces to
\begin{equation}
 E_w =\bar\kappa\oint
{\xi}^{A} D\varphi_A .
\label{witten}
\end{equation}
By choosing $\varphi^A_0= {\xi}^{A}$,
the energy expression (\ref{energy})
is related to the Witten expression $E_w$ (\ref{witten}),
while the spinors are now anti-commuting.

We can obtain a 4-component spinors formulation
by introducing a Majorana spinor
$\Psi=(\varphi_A,\overline{\varphi}^{A'})$ in the Weyl representation,
the spinor action
can be alternatively written as
\begin{equation}
 {\cal S}_{\Psi}[\Psi^\alpha, \omega^{ab} ]=
 \int{\cal L}_\Psi = 2
 \int  \overline{D\Psi} \wedge \gamma_5 D\Psi.
 \label{diracaction}
\end{equation}
The {\it real} SO(3,1) connection 1-form
$\omega^{ab}=\omega^{AB}\epsilon^{A'B'}+
\overline{\omega}^{A'B'}\epsilon^{AB}$
consists of the unprimed self-dual connection and its conjugate.

In the Weyl representation,
the covariant derivative of the Dirac spinor 1-forms is given by
\begin{equation}
\overline{D\Psi}=
(D\varphi^A\quad\overline{D}\overline{\varphi}_{A'}),
\qquad
D\Psi={D\varphi_A
\choose \overline{D}\overline{\varphi}^{A'}},
\end{equation}
and $\gamma_5:=\gamma_0\gamma_1\gamma_2\gamma_3=
{\rm i} \; {\rm diag}(-1,-1,1,1)$.
The Lagrangian in (\ref{diracaction}) can thus
be split into unprimed and primed parts:
\begin{equation}
 {\cal L}_\Psi
  = 2 {\rm i}\;D\varphi^A\wedge D\varphi_A
   -2 {\rm i}\;\overline{D}\overline{\varphi}^{A'}\wedge
     \overline{D}\overline{\varphi}_{A'} ,
\label{LPsi}
\end{equation}
where the second term is just the conjugate of the first term.
The action ${\cal S}_{\Psi}$ in (\ref{diracaction})
is therefore just the real part of the
action ${\cal S'}$ in (\ref{action1}).

Now a key observation is that ${\cal S}[z]$ is a holomorphic
function of $z$, where
$z:=(\varphi^{A},\omega^{AB})$.
Therefore, just as for analytic functions
of a finite number of variables, the derivative
 $\delta {\cal S}[z]/\delta z$ vanishes if and only if
the derivative of the real (or imaginary) part of
${\cal S}[z]$ vanishes. Thus, as far as the equations of motion
are concerned, the action ${\cal S}_{\Psi}$ (\ref{diracaction})
is {\it equivalent} to the original
chiral action ${\cal S}$.

In conclsion, we proposed a gauge theory for
the anticommuting
version of the quadratic spinor Lagrangian for general relativity.
The gauge potential is a connection of the Super SL(2,C) group.
The action is quadratic in the Super SL(2,C) curvature
and depends purely on gauge connection.
By breaking the symmetry of the Super SL(2,C)
topological gauge theory to SL(2,C),
a spinor metric is given and is related to
the spacetime metric.
In the Hamiltonian formulation,
the canonical momentum $\pi^i_{AB}$ conjugates to the SL(2,C) connection
$\omega_i^{AB}$
is related to the triad in Ashtekar's New Variables.
Because of the nature of using a Majorana spinor 1-form field,
$\pi^i_{AB}$ is allowed to be complex while keeping the
right degrees of freedom.
The reality condition is replaced by a Majorana condition.
This suggests a question: ``What are the local properties
of gravitational field?''
Among other properties, this letter suggests that the
Majorana property might be one of the local properties for gravitation,
which means that the resulting particle is
neutral so that it contains no charge.

\section*{Acknowledgements}

I would like to thank T.~Jacobson, L.~J.~Mason, J.~M.~Nester,
D.~C.~Robinson, R.~W.~Tucker and R.~M.~Wald for helpful comments;
the Mathematics Department, King's College London,
the Physics Department, National Central University,
Lancaster University,
and the Mathematical Institute at Oxford
for hospitality
while some of the work was carried out.
The research was supported in part by
NSC88-2112-M-008-018 (Taiwan) and EPSRC (UK).

\end{document}